\documentclass[epj]{svjour}

\def\be{\begin{equation}} 
\def\ee{\end{equation}} 

\def\bea{\begin{eqnarray}}
\def\ena{\end{eqnarray}}

\RequirePackage{graphicx}

\usepackage{cite} 
\usepackage{amsmath}
\usepackage{hyperref}
\usepackage{amsmath,amssymb}
\usepackage{xcolor}

\usepackage{graphics}

\begin{document}
\title{Quasinormal modes of regular black holes with Non-Linear-Electrodynamical sources}

\author{
Grigoris Panotopoulos \inst{1} 
\thanks{E-mail: \href{mailto:grigorios.panotopoulos@tecnico.ulisboa.pt}{\nolinkurl{grigorios.panotopoulos@tecnico.ulisboa.pt}} }
\and
\'Angel Rinc\'on \inst{2}
\thanks{E-mail: \href{mailto:angel.rincon@pucv.cl}{\nolinkurl{angel.rincon@pucv.cl}} }
}                     
%
%
\institute{ 
Centro de Astrof{\'i}sica e Gravita{\c c}{\~a}o, Departamento de F{\'i}sica, Instituto Superior T\'ecnico-IST,
\\
Universidade de Lisboa-UL, Av. Rovisco Pais, 1049-001 Lisboa, Portugal
\and
Instituto de F{\'i}sica, Pontificia Universidad Cat\'olica de Valpara{\'i}so, Avenida Brasil 2950, Casilla 4059, Valpara{\'i}so, Chile
}

\date{Received: date / Revised version: date}
%
\abstract{
We compute the spectrum of quasinormal frequencies of regular black holes obtained in the presence of Non-Linear Electrodynamics. In particular, we perturb the black hole with a minimally coupled massive scalar field, and we study the corresponding perturbations adopting the 6th order WKB approximation. We analyze in detail the impact on the spectrum of the charge of the black hole, the quantum number of angular momentum and the overtone number. All modes are found to be stable. Finally, a comparison with other charged black holes is made, and an analytical expression for the quasinormal spectrum in the eikonal limit is provided.
}

\maketitle

\section{Introduction}

Although black holes (BHs) and gravitational waves (GWs) are predicted to exist within the framework of Einstein's General Relativity (GR) \cite{GR}, until a few years ago there was only indirect evidence for the existence of both of them. Galactic centres are supposed to host supermassive BHs \cite{SMBH1,SMBH2,SMBH3}, while gravitational waves had been indirectly seen in orbital decay of binary systems due to emission of gravitational radiation \cite{Binary}. The historical LIGO direct detection of GWs \cite{ligo1,ligo2,ligo3} has provided the strongest evidence so far that BHs do exist in Nature and that they merge, and it has opened a completely new window to the Universe. Despite the fact that BHs are the simplest objects in the Universe, characterized entirely by a handful of parameters, such as mass, spin and charges, they are fascinating objects bringing together many different areas, from gravitation to thermodynamics to quantum mechanics, and they are of paramount importance to gravitation, since they have the potential of revealing (at least) some of the hidden secrets of quantum gravity.

A special attention is devoted to non-linear electrodynamics (NLE), which has a long history and it has been studied over the years in several different contexts. Maxwell's classical theory is based on a system of linear equations, but when quantum effects are taken into account, the effective equations become non-linear. The first models go back to the 30's when Euler and Heisenberg obtained QED corrections \cite{Euler}, while Born and Infeld obtained a finite self-energy of point-like charges \cite{BI}. Furthermore, a 
non-trivial extension of Maxwell's theory leads to the by now well-known
Einstein-power-Maxwell (EpM) theory described by a Lagrangian density of the form $\mathcal{L}(F) = F^k$, where $F$ is the usual Maxwell invariant,
to be defined below, and $k$ is an arbitrary rational number. This class of theories maintain the nice properties of conformal invariance in any number of space time dimensionality $D$ provided that $k=D/4$. Black hole solutions in (1+2)-dimensional and higher-dimensional EpM theories have been obtained in \cite{BH1} and \cite{BH2} respectively (see also \cite{SD1,SD2} for scale-dependent black holes in the presence of EpM NLE). More exotic NLE models, such as logarithmic \cite{Soleng:1995kn}, rational \cite{Kruglov_2014}, and exponential \cite{Hendi:2013dwa} among others, have been studied in connection to black hole physics.

Furthermore, the well-known Reissner-Nordstr{\"o}m black hole solution \cite{RN} is a charged solution to Einstein's field equations coupled to Maxwell's linear electrodynamics, and it is characterized by a singularity at the center. The singularity is hidden by an event horizon, and therefore it has no effect on the outside region, where Physics is well-behaved. The existence of singularities, however, indicate the breakdown of General Relativity, and so attempts are made to obtain regular BH solutions, such as the solution obtained for the first time by Bardeen \cite{Bardeen1}, see also \cite{borde,Bardeen2}. In \cite{borde} this class of black holes was named ``Bardeen black holes", while in \cite{Bardeen2} it was shown that the Bardeen model could be obtained within Einstein's General Relativity coupled to an appropriate non-linear electromagnetic source. Nowadays several regular black hole solutions are known, which have been obtained applying the same approach, that is assuming appropriate non-linear electromagnetic sources, which in the weak field limit are reduced to the standard Maxwell's linear theory. It is an approach that allows us to generate a new class of solutions to Einstein's field equations \cite{beato1,beato2,beato3,bronnikov,dymnikova,hayward,vagenas1,vagenas2,culetu}, which on the one hand have a horizon, and on the other hand their curvature invariants, such as the Ricci scalar $R$, are regular everywhere, as opposed to the standard Reissner-Nordstr{\"o}m solution. Regular BHs may help us understand the final states of gravitational collapse \cite{ellis,extra}, which is not possible when singularities are present.

Quasinormal modes (QNMs) are complex numbers that encode the information on how the black holes relax after the perturbation is applied. They depend on the type of perturbation (scalar, vector etc) and on the properties of the geometry itself, and they do no depend on initial conditions. The work of \cite{wheeler} marked the birth of BH perturbations, and it was later extended by \cite{zerilli1,zerilli2,zerilli3,moncrief,teukolsky}. A comprehensive overview of BH perturbations is summarized in Chandrasekhar's monograph \cite{monograph}. Black hole perturbation theory and QNMs are relevant during the ringdown phase, in which a distorted object after the merging of two black holes is formed, and the geometry of spacetime undergoes dumped oscillations due to the emission of gravitational waves. For a review on the subject see \cite{review1}, and for a more recent ones \cite{review2,review3}.

Given the interest in gravitational wave Astronomy and on QNMs of black holes, it would be interesting to see what kind of QN spectra are expected from regular BHs. In previous works quasinormal modes of regular black holes were computed by several authors, see e.g. \cite{correa,lemos,otro,wu,ahmedov1,ahmedov2,bambi,ahmedov3}. It is the goal of the present article to compute the QNMs of regular charged black holes in the presence of non-linear electrodynamics assuming a massive scalar field as the test particle that perturbs the black hole. In particular, we shall consider one of the black holes obtained in \cite{vagenas2} employing mass distribution functions. This regular black hole has already been discussed in \cite{wu}, our work however is different in several respects. In particular, i) we study scalar instead of gravitational perturbations, ii) we compute the QNMs for more values of the charge of the black hole approaching extremality, and iii) we adopt the WKB method of sixth order instead of third.

The plan of our work is the following: After this Introduction, we present the theory and the regular BH solution in subsection 2.1, and the wave equation for scalar perturbations in subsection \ref{WE_QNM}. In the third Section we compute the QNMs of the black holes in the WKB approximation and we discuss our results. Finally, we conclude our work in Section \ref{Conclusions}. We use natural units such that $c = G = \hbar = 1$ and metric signature $(-, +, +, +)$.

\section{Formalism}

\subsection{Regular black hole in the presence of NLE}

Let us consider a 4-dimensional theory described by the action
\begin{equation}
S[g_{\mu \nu}, A_\mu] = \int \mathrm{d} ^4x \sqrt{-g} \left[ \frac{1}{2 \kappa} R - \frac{1}{4 \pi} \: \mathcal{L}(F) \right],
\end{equation}
where $R$ is the Ricci scalar, $g$ the determinant of the metric tensor $g_{\mu \nu}$, $\kappa=8 \pi$, and $F \equiv (1/4) F_{\mu \nu} F^{\mu \nu}$ the Maxwell invariant with $F_{\mu \nu}$ being the electromagnetic field strength. 

Varying the action with respect to the Maxwell potential $A_\mu$ one obtains the generalized Maxwell equations \cite{vagenas2}
\begin{equation}
\partial_\mu (\sqrt{-g} F^{\mu \nu} \mathcal{L}_F) = 0.
\end{equation}
where we define $\mathcal{L}_F \equiv d \mathcal{L}/dF$. Furthermore, varying the action with respect to the metric tensor one obtains Einstein's field equations
\begin{equation}
G_{\mu \nu}  = 8 \pi T_{\mu \nu}
\end{equation}
where $G_{\mu \nu}$ is the Einstein tensor, while the matter stress-energy tensor $T_{\mu \nu}$ corresponds to the electromagnetic energy-momentum tensor \cite{vagenas2}
\begin{equation}
T_{\mu \nu} = \mathcal{L}(F) g_{\mu \nu} -\mathcal{L}_F F_{\mu \sigma} F_{\nu}^{\sigma}
\end{equation}
We seek spherically symmetric static solutions of the form
\begin{equation}
\mathrm{d}s^2 = -f(r) \mathrm{d}t^2 + f(r)^{-1} \mathrm{d}r^2 + r^2 (\mathrm{d} \theta^2 + \sin^2 \theta \mathrm{d} \varphi^2)
\end{equation}
but instead of specifying the Lagrangian density $\mathcal{L}$ to obtain the metric function $f(r)$, we assume that the solution for the metric lapse function corresponds to a known function instead \cite{vagenas2}
\begin{equation}
f(r) = 1 - \frac{2 M}{r} \: \exp\left( - \frac{q^2}{2 M r}  \right)
\end{equation}
where $M,q$ are the mass and the electric charge of the black hole, respectively. Among several possible choices we have opted for this particular regular black hole since the exponential function is a common place in Physics. A couple of notable examples are for instance the Liouville-type potential for the dilaton \cite{expo1} or the Yukawa potential in nuclear physics that inspired the study of Yukawa black holes \cite{expo2}.

Introducing the distribution function $\sigma(r) = \exp(-q^2/(2Mr))$, the temporal and radial components of Einstein's equations read
\begin{equation}
- \frac{2 M}{r^2} \sigma'(r) = 8 \pi (\mathcal{L}(F)+E^2 \mathcal{L}_F)
\end{equation}
while the angular components read
\begin{equation}
- \frac{M}{r} \sigma''(r) = 8 \pi \mathcal{L}(F)
\end{equation}
Finally, the generalized Maxwell's equations are equivalent to the following equation
\begin{equation}
E \mathcal{L}_F  = - \frac{q}{4 \pi r^2}
\end{equation}
where the prime denotes differentiation with respect to $r$, and $E=F_{tr}$ is the electric field. Combining the equations of motion one can determine the electric field $E(r)$ as well as the electromagnetic Lagrangian density as functions of the radial coordinate
\begin{eqnarray}
E(r) & = & - \frac{M r^3}{2 q} \frac{\mathrm{d}}{\mathrm{d}r} \left( \frac{\sigma'(r)}{r^2} \right) \\
\mathcal{L}(r) & = & - \frac{M \sigma''(r)}{8 \pi r}
\end{eqnarray}
and so one obtains the electromagnetic Lagrangian $\mathcal{L}(F)$ in parametric form $F(r)=-(1/2)(E(r))^2$, $\mathcal{L}(r)$. Finally, the Ricci scalar is computed to be
\begin{equation}
R  = \frac{2 M}{r^2} (2 \sigma'(r) + r \sigma''(r))
\end{equation}
and it is regular everywhere.

\subsection{Perturbations for a test massive scalar field}\label{WE_QNM}
First we consider a four-dimensional spherically symmetric fixed gravitational background of the form
\begin{equation}
\mathrm{d}s^2 = -f(r) \mathrm{d}t^2 + f(r)^{-1} \mathrm{d}r^2 + r^2 (\mathrm{d} \theta^2 + \sin^2 \theta \mathrm{d} \phi^2)
\end{equation}
where the metric function is given by \cite{vagenas2}
\begin{equation}
f(r) = 1-\frac{2 M}{r} \exp \left(-\frac{q^2}{2Mr}\right)
\end{equation}
with $M$ and $q$ being the mass and the electric charge of the black hole, respectively.
Clearly, when $q=0$ we recover the Schwarzschild solution for a neutral BH, while expanding in powers of $q$ we obtain the following approximate expression
\begin{equation}
f(r) \approx 1-\frac{2 M}{r} + \frac{q^2}{r^2} - \frac{q^4}{4 M r^3} + \mathcal{O}(q^6)
\end{equation}
The first 3 terms are precisely the ones corresponding to the well--known RN black hole solution, while higher powers are corrections to that.

\begin{figure*}[ht]
\centering
\includegraphics[width=0.32\textwidth]{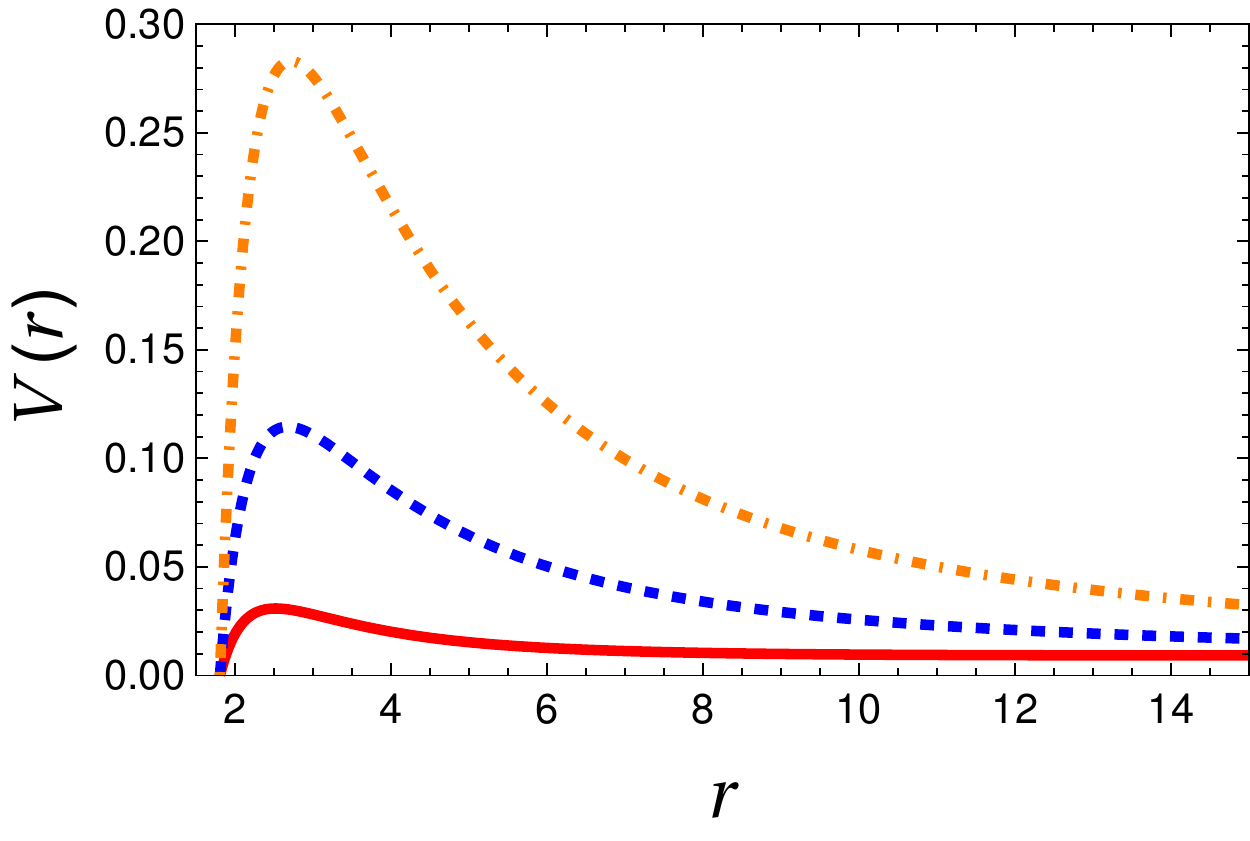}   \
\includegraphics[width=0.32\textwidth]{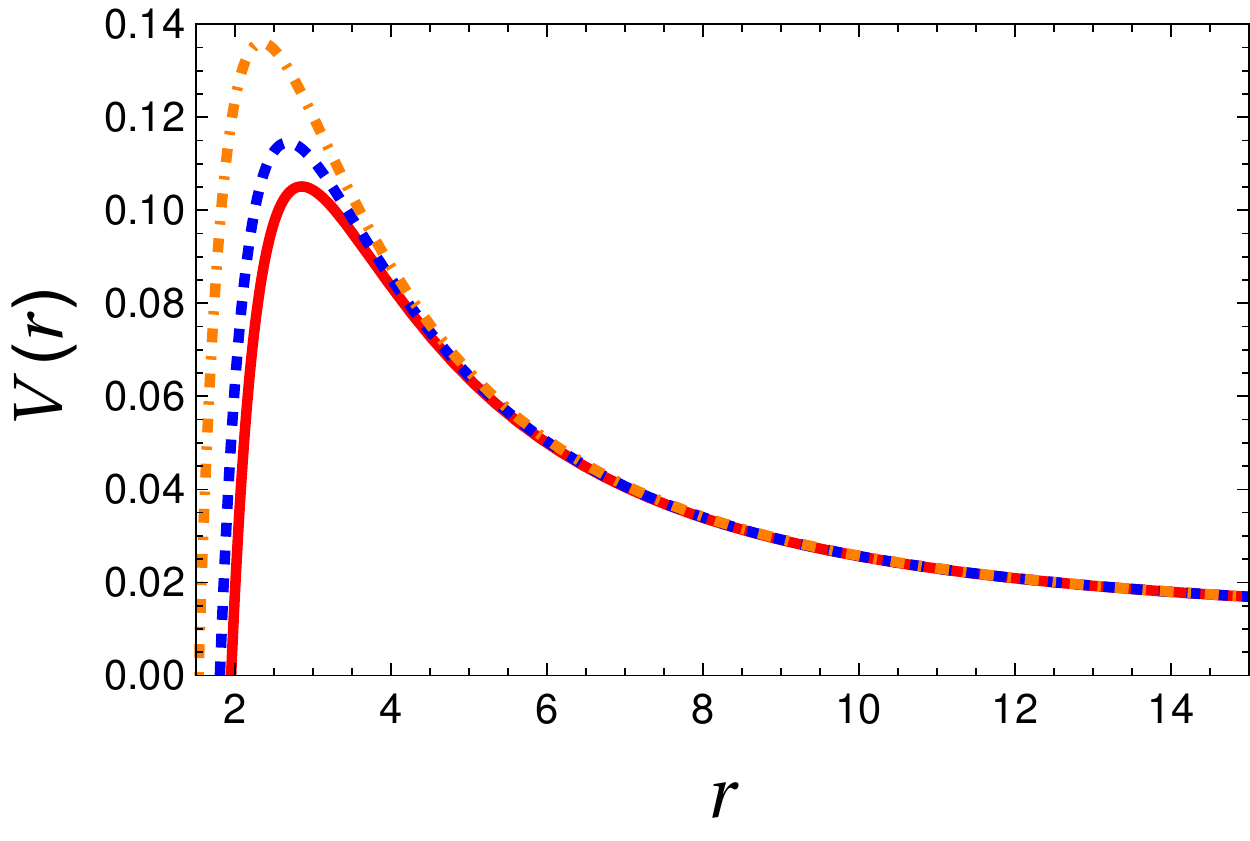}   \
\includegraphics[width=0.32\textwidth]{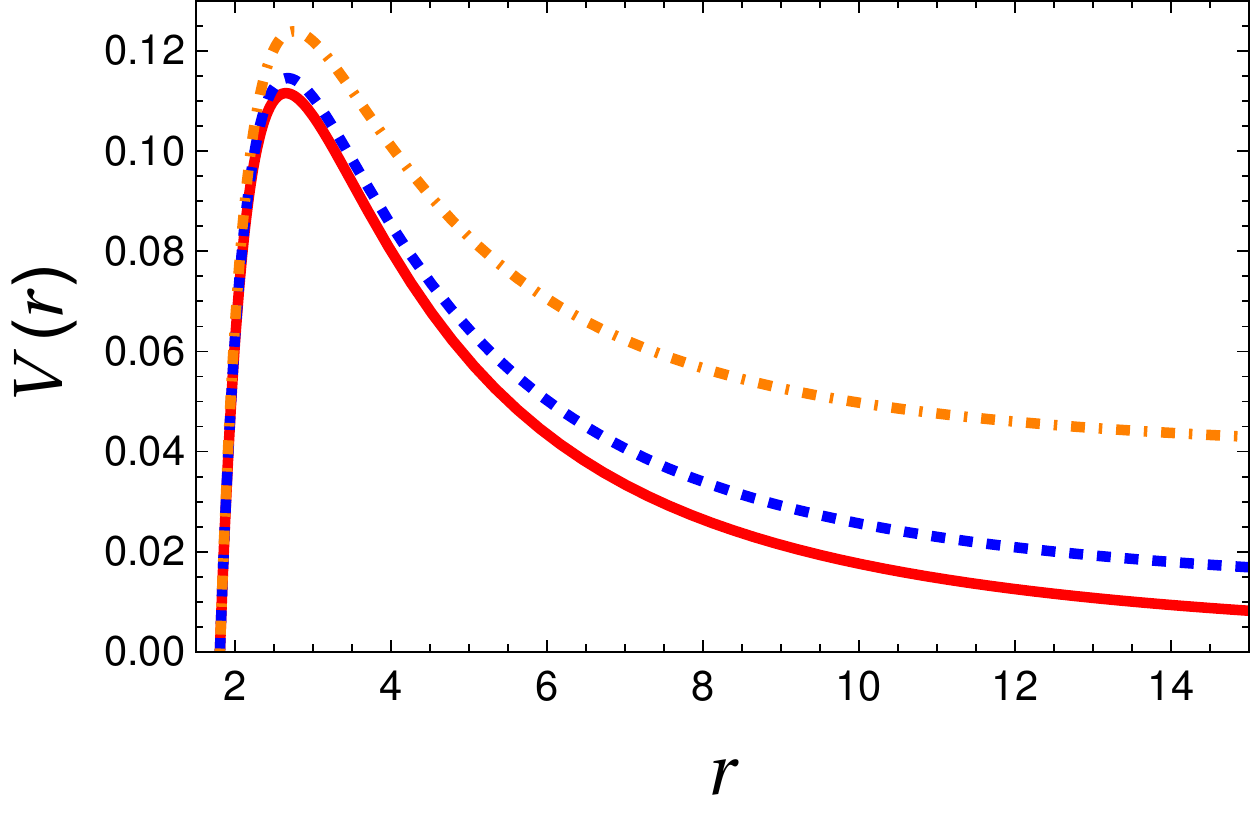}   \
\caption{
The panels in the first (left), second (center) and third (right) show $V(r)$ for: 
1) Effective potential $V(r)$ as a function of the radial coordinate $r$ assuming $M=1$, $q=0.6$ and $\mu=0.1$ for three different cases: 
i)   $l=0$ (solid red line), 
ii)  $l=1$ (dashed blue line) and 
iii) $l=2$ (dotted--dashed orange line). 
2) Effective potential $V(r)$ as a function of the radial coordinate $r$ assuming $M=1$, $l=1$ and $\mu=0.1$ for three different cases: 
i)   $q=0.3$ (solid red line), 
ii)  $q=0.6$ (dashed blue line) and 
iii) $q=0.9$ (dotted--dashed orange line).
3) Effective potential $V(r)$ as a function of the radial coordinate $r$ assuming $M=1$, $q=0.6$ and $l=1$ for three different cases: 
i)   $\mu=0$ (solid red line), 
ii)  $\mu=0.1$ (dashed blue line) and 
iii) $\mu=0.2$ (dotted--dashed orange line).
}
\label{fig:potential}
\end{figure*}

Then, we perturb the black hole with a probe minimally coupled massive scalar field with equation of motion
\begin{equation}
\frac{1}{\sqrt{-g}} \partial_\mu (\sqrt{-g} g^{\mu \nu} \partial_\nu) \Phi = \mu^2 \Phi
\end{equation}
where $\mu$ is the mass of the test scalar field. We separate variables making the standard ansatz
\begin{equation}\label{separable}
\Phi(t,r,\theta, \phi) = e^{-i \omega t} \frac{\psi(r)}{r} Y_l^m (\theta, \phi)
\end{equation}
with $Y_l^m$ being the spherical harmonics, and we obtain a Schr{\"o}dinger-like equation of the form
\begin{equation}
\frac{\mathrm{d}^2 \psi}{\mathrm{d}x^2} + (\omega^2 - V(x)) \psi = 0
\end{equation}
with $x$ being the so--called tortoise coordinate
\begin{equation}
x  =  \int \frac{\mathrm{d}r}{f(r)}
\end{equation}
while the effective potential is given by the expression
\begin{equation}
V(r) = f(r) \: \left( \mu^2 + \frac{l (l+1)}{r^2}+\frac{f'(r)}{r} \right)
\end{equation}
where the prime denotes differentiation with respect to $r$.
The effective potential as a function of the radial coordinate can be seen in Figure \eqref{fig:potential}. The left panel shows the impact of angular momentum, the right panel shows the impact of the mass of the test scalar field that perturbs the black hole, while the panel in the middle shows the impact of the electric charge of the black hole.

To complete the formulation of the physical problem at hand we must add the appropriate boundary conditions, both at infinity and at the horizon, to the Schr{\"o}dinger-like equation. For asymptotically flat spacetimes we impose the quasinormal condition \cite{valeria}.
\begin{equation}
\psi(x) \rightarrow
\left\{
\begin{array}{lcl}
A e^{-i \omega x} & \mbox{ if } & x \rightarrow - \infty \\
&
&
\\
 C e^{i \omega x}  & \mbox{ if } & x \rightarrow + \infty
\end{array}
\right.
\end{equation}
where the parameters $A,C$ are arbitrary constants. The physical meaning of the required boundary conditions is the following: i) on the one hand the purely ingoing wave expresses the fact that nothing escapes from the horizon, and ii) on the other hand the purely outgoing wave corresponds to the requirement that no radiation comes from infinity \cite{valeria}.
It is precisely the quasinormal condition that allows us to obtain an infinite set of discrete complex numbers, which are the so-called quasinormal frequencies of the black hole. 

Given the time dependence of the scalar field, $\sim e^{-i \omega t}$, the mode 
grows exponentially (in other words it is unstable) when $\omega_I > 0$, or decays exponentially (it is stable) when $\omega_I < 0$. In the latter case the real part determines the frequency of the oscillation, $\omega_R/(2 \pi)$, while the inverse of $|\omega_I|$ determines the dumping time, $t_D^{-1}=|\omega_I|$.

\begin{figure*}[ht]
\centering
\includegraphics[width=0.49\textwidth]{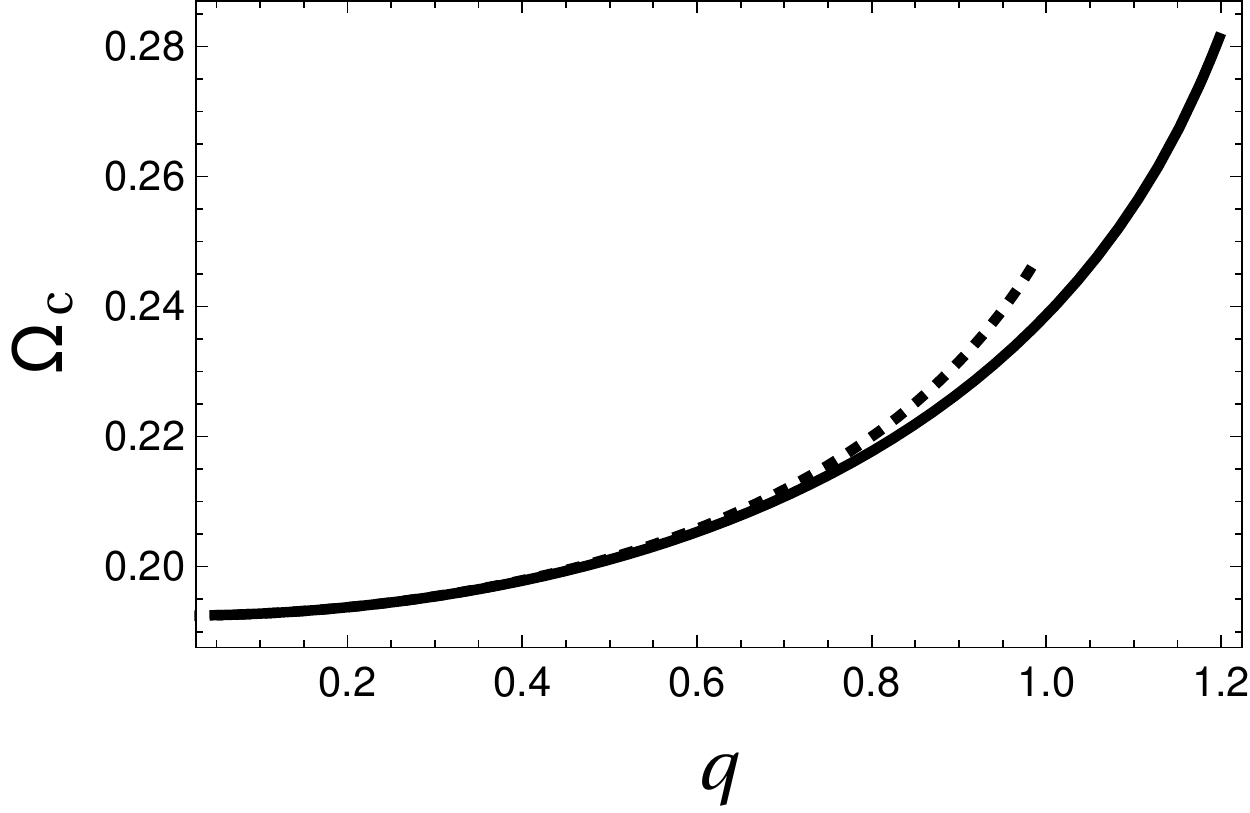}   \
\includegraphics[width=0.49\textwidth]{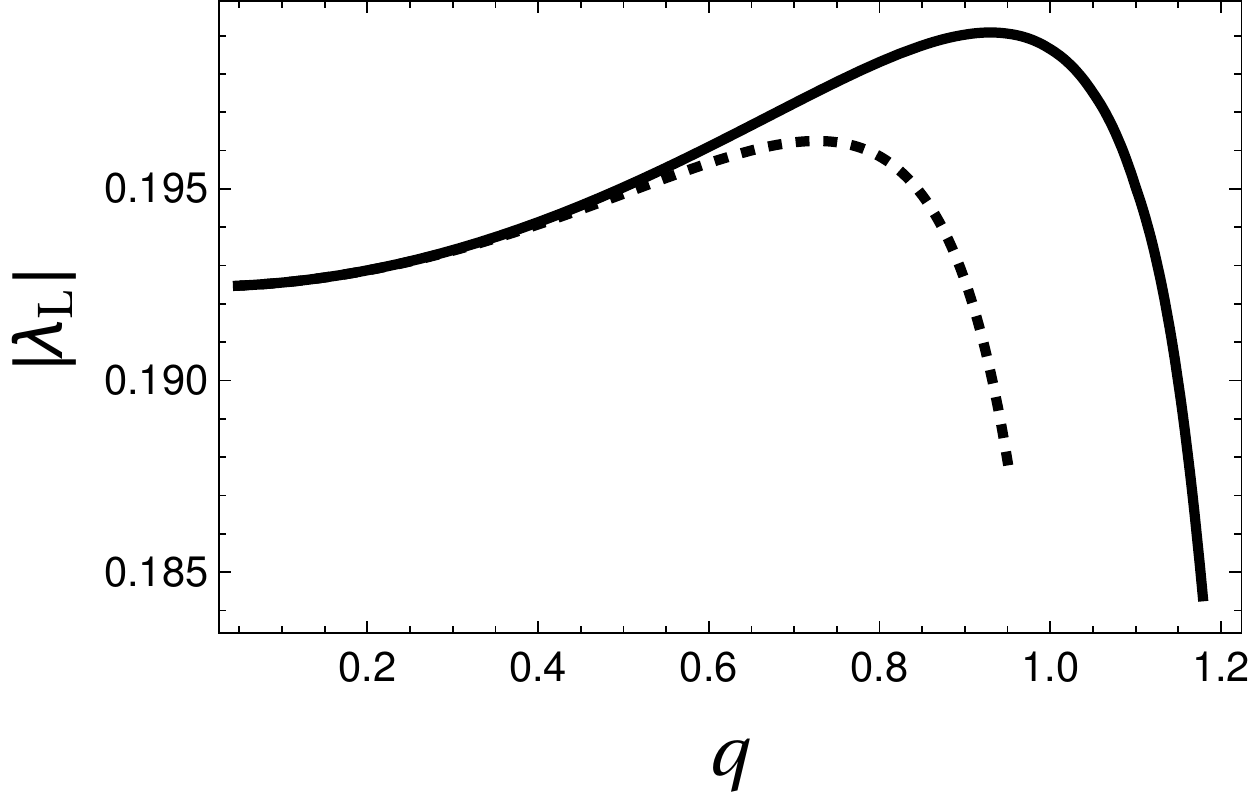}   \
\caption{
QNMs in the eikonal limit: Angular velocity (left panel) and Lyapunov exponent (right panel) vs electric charge for $M=1$.
The solid curves corresponds to the regular BH, while the dashed curves to the RN solution. 
}
\label{fig:eikonal}
\end{figure*}

\section{QNMs of regular BHs in the WKB approximation}

\subsection{Numerical results}

An exact computation of the QNMs of black holes in an analytical way is possible only in some cases, see e.g. \cite{potential,ferrari,cardoso2,exact1,exact2,exact3,exact4,exact5,exact6,Ovgun:2018gwt,Rincon:2018ktz}. Semi-analytical approaches based on the popular WKB approximation \cite{wkb1,wkb2} (for an improved semi-analytical approach see \cite{improved}) have been extensively applied to several cases. For previous similar works see e.g. \cite{paper1,paper2,paper3,paper4,paper5,paper6}, and for more recent works see e.g. \cite{correa,paper8,paper9,paper10,Rincon:2018sgd}, and references therein. 

The QN frequencies may be computed using the formula
\begin{equation}
\omega^2 = V_0+(-2V_0'')^{1/2} \Lambda(n) - i \nu (-2V_0'')^{1/2} [1+\Omega(n)]
\end{equation}
where $V_0$ is the maximum of the effective potential, $V_0''$ is its second derivative evaluated at the maximum, $n=0,1,2...$ is the so-called overtone number, $\nu=n+1/2$, while $\Lambda(n), \Omega(n)$ are 
non-trivial expressions of $\nu$ and higher derivatives of the potential evaluated at the maximum, and can be seen e.g. in \cite{correa,paper2}. 

In the present work we have made use of the Wolfram Mathematica \cite{wolfram} code with WKB at any order from one to six (here we have worked in 6th order) presented in \cite{code}. We have fixed the mass of the black hole to be $M=1$, the mass of the test scalar field is taken to be either $\mu=0$ or $\mu=0.1$, while for the electric charge $q$ we have considered 6 values from 0.2 to 1.2 (the extremal black hole is obtained for $q \simeq 1.213$), so that the numerical results go smoothly to the Schwarzschild limit as $q \rightarrow 0$. In \cite{cardoso3} it was shown that the WKB approximation works very well for $l > n$, and therefore here we shall consider the cases i) ${l=1,n=0}$, ii) ${l=2,n=0,n=1}$ and iii) ${l=3,n=0,n=1,n=2}$. We have also included the lowest multiple ${l=0,n=0}$ since it corresponds to the fundamental mode, although we do not expect the approximation to be very good. Finally, we consider non-extremal black holes (for the extremal case see \cite{extremal}), while the eikonal regime $l \gg 1$ will be considered separately in the end before concluding our work.

Our numerical results for the QN modes of the charged regular black holes are summarized in the tables below. In the last 2 tables we have made a direct comparison between regular black holes and the standard Reissner-Nordstr{\"o}m (RN) black hole \cite{RN} for $l=1,n=0$ and $l=2,n=0$. The values of the frequencies corresponding to the massless scalar field case are shown in parenthesis. We see that for $\mu=0$ the real part is lower, while the imaginary part is more negative compared to the $\mu=0.1$ case. 

\begin{figure*}[ht]
\centering
\includegraphics[width=0.49\textwidth]{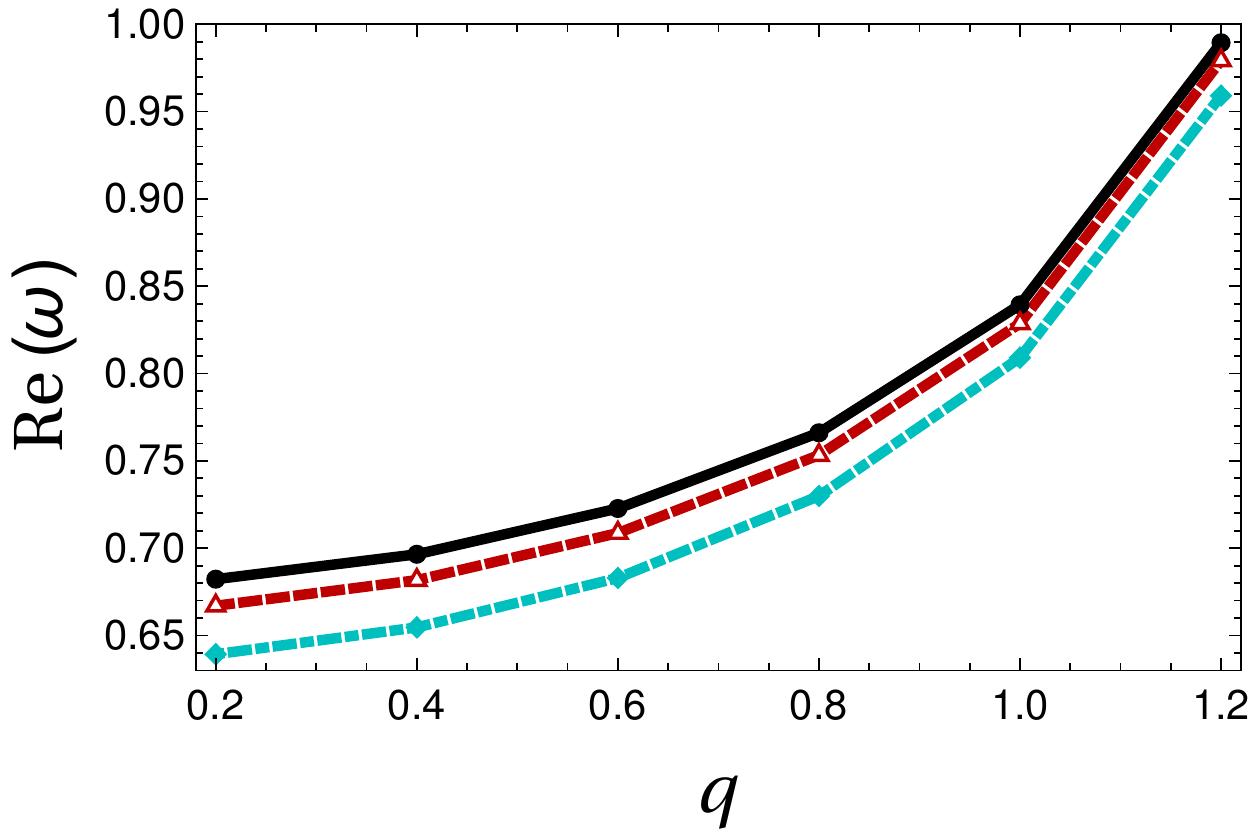}   \
\includegraphics[width=0.49\textwidth]{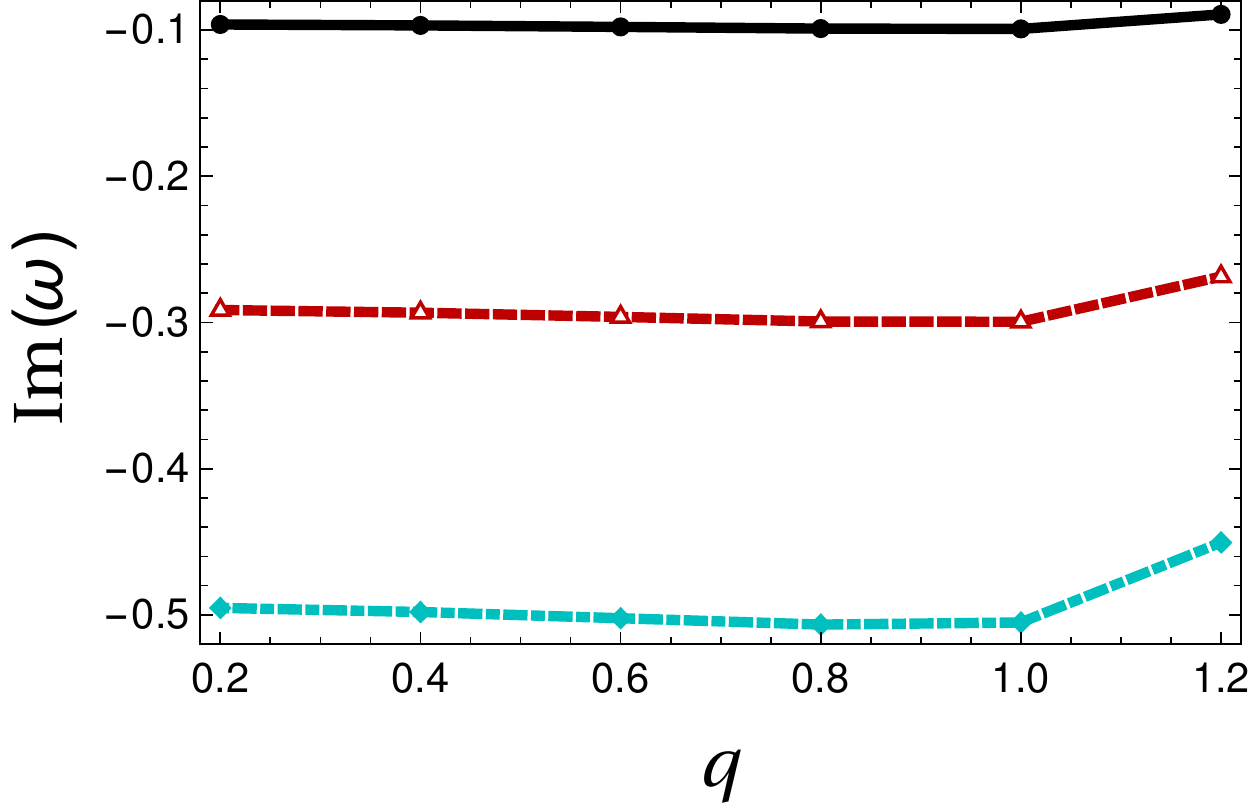}   
\\
\includegraphics[width=0.49\textwidth]{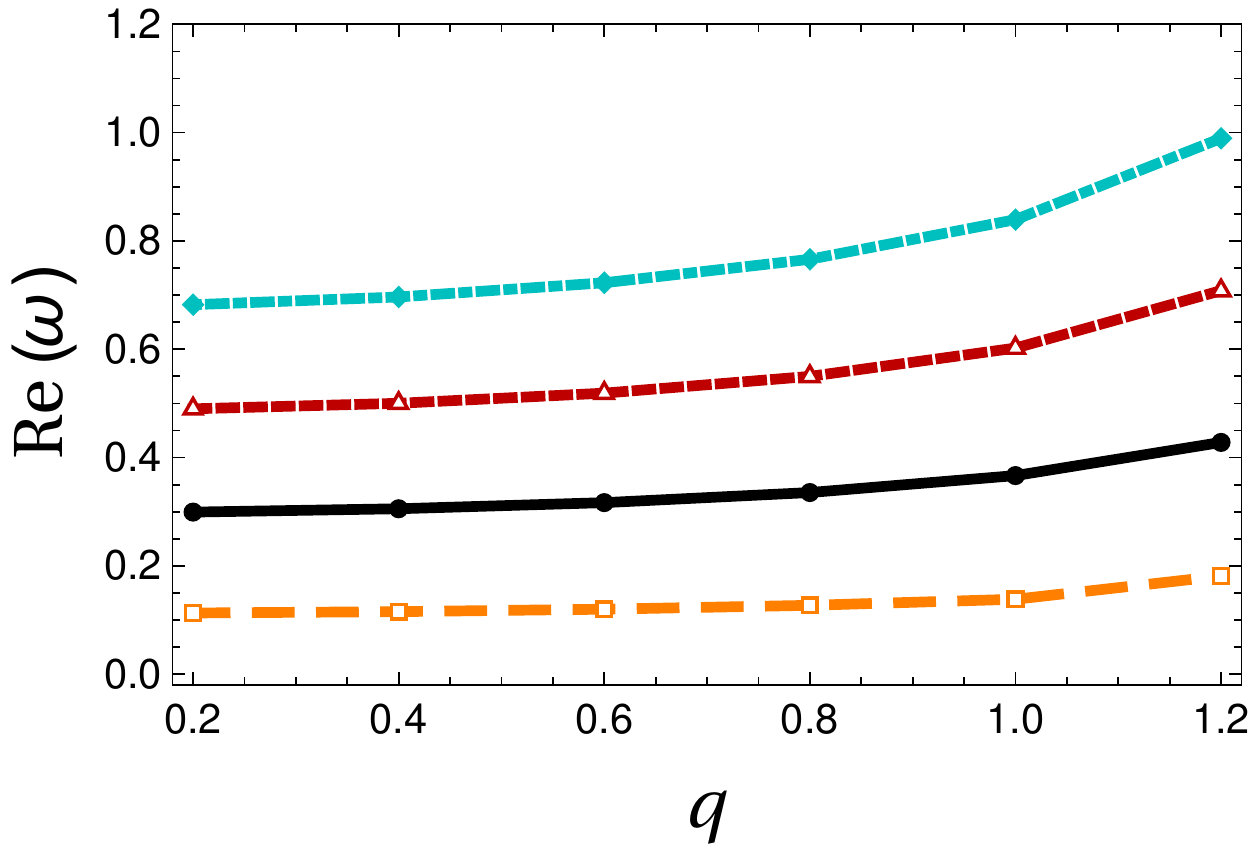}   \
\includegraphics[width=0.49\textwidth]{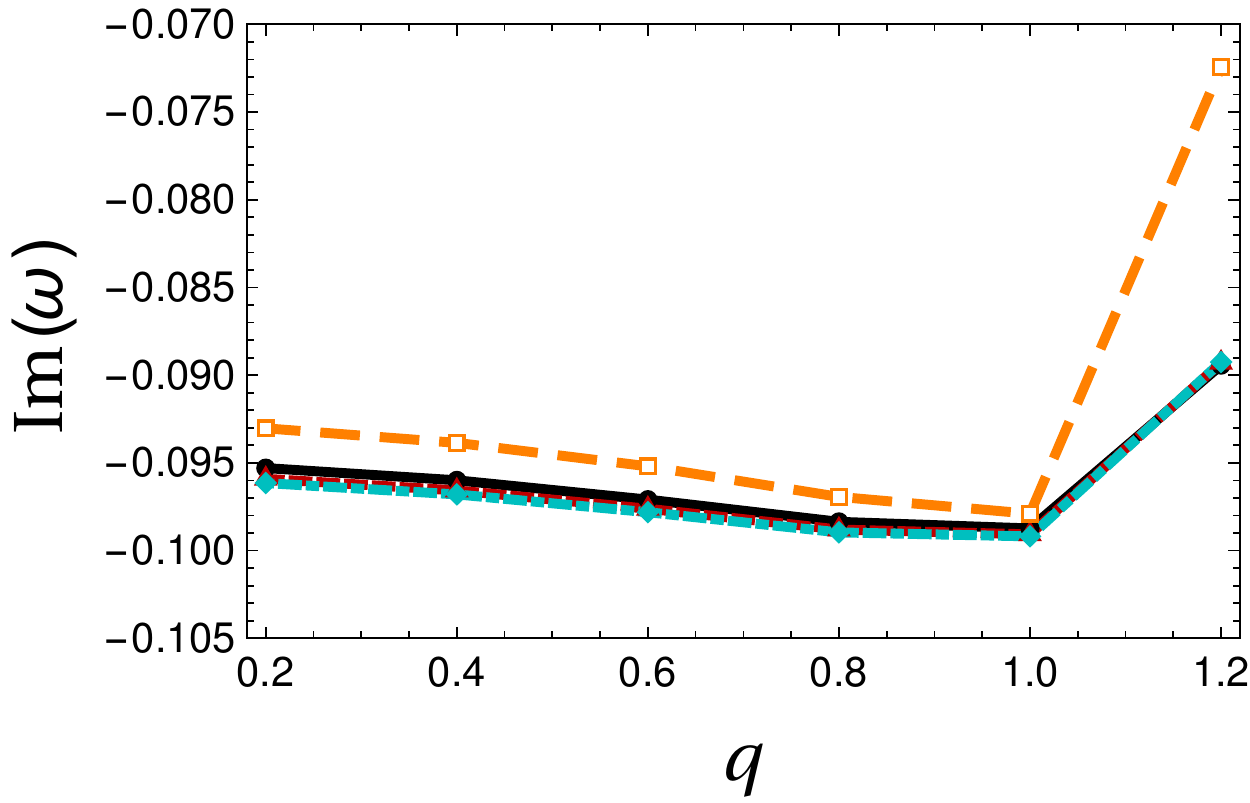} 
\caption{
The first row shows the real and imaginary quasinormal frequencies vs the electric charge of the black hole taking $l=3$, $M=1$ and $\mu = 0.1$ for: 
i)   $n=0$ (solid black line),
ii)  $n=1$ (dashed red line) and
iii) $n=2$ (dotted--dashed cyan line).
The second row shows the real and imaginary quasinormal frequencies vs the electric charge of the black hole taking $n=0$, $M=1$ and $\mu = 0.1$ for: 
i)   $l=1$ (solid black line),
ii)  $l=2$ (dashed red line),
iii) $l=3$ (dotted--dashed cyan line) and
iv)  $l=0$ (long dashed orange line).
}
\label{fig:modes}
\end{figure*}

For better visualization we show graphically how the QNMs depend on the angular momentum, the overtone number as well as the electromagnetic coupling. In particular, the first row of \ref{fig:modes} shows the real and imaginary part of the frequencies versus $q$ for 3 different values of the overtone number $n=0,1,2$ and fixed angular momentum $l=3$, while the second row of the same figure shows the real and imaginary part of the frequencies versus $q$ for 4 different values of angular momentum $l=0,1,2,3$ and fixed $n=0$. The real part increases with the electric charge and with the angular momentum, like in the case of RN. In other theories too, such as Born-Infeld and Gauss-Bonnet gravity, the real part of QNM of charged BH increases with the electric charge \cite{paper4,paper5}. The imaginary part becomes more and more negative with the electric charge, and less and less negative with the angular momentum, like in the RN case \cite{RN2}. Notice that the characteristic minimum of the imaginary part (or maximum if one plots $-\text{Im}(\omega)$ versus $q$) for a certain value of the electric charge close to its extremal value is observed, like in the RN case (also observed in \cite{paper10} for charged BHs in the EpM theory) and contrary to Born-Infeld NLE and Gauss-Bonnet gravity, where the imaginary part is a monotonic function of the electric charge \cite{paper4,paper5}.

\subsection{QNMs in the eikonal limit}

In the eikonal regime ($l \gg 1$) the WKB approximation becomes increasingly accurate, and therefore one can obtain analytical expressions for the QN frequencies. In the eikonal limit ($l \rightarrow \infty$) it is the angular momentum term that dominates in the expression for the effective potential
\begin{equation}
V(r) \approx \frac{f(r) l^2}{r^2} \equiv l^2 g(r)
\end{equation}
where for convenience we have introduced a new function $g(r)=f(r)/r^2$. It is not difficult to verify that the maximum of the potential is located at the point $r_1$, which may be computed solving the following algebraic equation
\begin{equation}
2 f(r_1) - r_1 f'(r)|_{r_1} = 0
\end{equation}
Then, following the formalism developed in \cite{eikonal1}, the QNMs in the eikonal regime are computed by
\begin{equation}
\omega_{l \gg 1} = \Omega_c l - i \left(n+\frac{1}{2}\right) |\lambda_L|
\end{equation}
where the Lyapunov exponent $\lambda_L$ is given by \cite{eikonal1}
\begin{equation}
\lambda_L = r_1^2 \sqrt{\frac{g''(r_1) g(r_1)}{2}}
\end{equation}
while the angular velocity $\Omega_c$ at the unstable null geodesic is given by \cite{eikonal1}
\begin{equation}
\Omega_c = \frac{\sqrt{f(r_1)}}{r_1}
\end{equation}
Notice that the same expressions for $\Omega_c$ and for $\lambda_L$ may be obtained applying the WKB approximation of 1st order, as it was done e.g. in \cite{Ponglertsakul:2018smo}. Therefore, although in the presence of non-linear electromagnetic sources photons follow the null trajectories of an effective geometry rather than the null geodesics of the true geometry \cite{eikonal2,eikonal3,ref2}, the previous expressions for $\Omega_c$ and for $\lambda_L$ remain the same.
We see that the angular velocity determines the real part of the modes, where only the degree of angular momentum $l$ enters, while the Lyapunov exponent determines the imaginary part of the modes, where only the overtone number $n$ enters. In Fig. \eqref{fig:eikonal} we show the angular velocity (left panel) as well as the Lyapunov exponent (right panel) as a function of $q$ for $M=1$. For comparison reasons we show in the same figure the corresponding quantities for the standard RN solution. The angular velocity increases monotonically with the electric charge, while the Lyapunov exponent reaches a maximum value first and then it decreases as opposed to the Bardeen black hole studied in \cite{correa}, where it was found that $\lambda_L$ decreases monotonically with $q$. The angular velocity of the regular BH lies below $\Omega_c$ of the RN BH, while the Lyapunov exponent of the regular BH lies above $\lambda_L$ of the RN solution. The figure show that for the same BH parameters $M,q$ and the same angular degree $l$, the regular BH solution is characterized by a lower real part and a higher imaginary part, and the differences grow as we approach extremality.

Finally, when studying the QNMs of a black hole stability of the system is one of the most important results. The conditions obtained in \cite{Moreno}
allow us to make a simple test on the dynamical stability of the black hole studied here. The conditions are the following
\begin{eqnarray}
\mathcal{H} & < & 0 \\
\mathcal{H}_x & < & 0 \\
\mathcal{H}_{xx} & < & 0 \\
3 \mathcal{H}_x & \leq & x f \mathcal{H}_{xx}
\end{eqnarray}
where $\mathcal{H}$ is the Hamiltonian density of the system, and $x=q^2/r^2$. These conditions should hold for any $r > r_H$ or in the range $0 < x < x_H$, where $x_H=(q/r_H)^2$. For the exponential mass distribution function considered in this work the Hamiltonian is computed to be  \cite{vagenas2}
\begin{equation}
\mathcal{H} = P e^{-U}
\end{equation}
where $P=(1/4) P_{\mu \nu} P^{\mu \nu}$, $P_{\mu \nu}=\mathcal{L}_F F_{\mu \nu}$, and $U$ is given by
\begin{equation}
U = \left( \frac{q}{2 M} \right) (-2 q^2 P)^{1/4}
\end{equation}
The Hamiltonian density as a function of $x$ is found to be
\begin{equation}
\mathcal{H}  =  - \frac{x^2}{32 \pi^2 q^2} \exp\left(- \frac{q \sqrt{x}}{4 M \sqrt{\pi}} \right)
\end{equation}
and its derivatives can be computed in a straightforward manner. It is easy to verify that all the aforementioned conditions are satisfied for $M=1$ and
$0 < q < 1.2$.

\begin{table*}
\centering
\caption{QN frequencies for $q=0.20,M=1,\mu=0.1$. Massless case in parenthesis}
\begin{tabular}{ccccc}
\hline
$n$ & $l=0$ & $l=1$ & $l=2$ & $l=3$ \\
\hline
0  & 0.1130-0.0930 i & 0.2993-0.0953 i & 0.4900-0.0959 i & 0.6823-0.0961 i \\
   & (0.1113-0.1010 i) & (0.2949-0.0980 i) & (0.4869-0.0970 i) & (0.6799-0.0967 i) \\
\hline
1 &  &  & 0.4689-0.2939 i & 0.6670-0.2915 i \\
  &  &  & (0.4673-0.2962 i) & (0.6653-0.2929 i) \\
\hline
2 &  &  &  & 0.6393-0.4952 i \\
  &  &  &  & (0.6385-0.4970 i) \\ 
\hline
\end{tabular}
\label{table:First_set}
\end{table*}

\begin{table*}
\centering
\caption{QN frequencies for $q=0.40,M=1,\mu=0.1$. Massless case in parenthesis}
\begin{tabular}{ccccc}
\hline
$n$ & $l=0$ & $l=1$ & $l=2$ & $l=3$ \\
\hline
0  & 0.1155-0.0939 i & 0.3055-0.0960 i & 0.5002-0.0966 i & 0.6965-0.0968 i  \\
   & (0.1137-0.1016 i) & (0.3012-0.0985 i) & (0.4972-0.0976 i) & (0.6943-0.0973 i)  \\
\hline   
1 &  &  & 0.4796-0.2957 i & 0.6817-0.2933 i  \\
  &  &  & (0.4781-0.2979 i) & (0.6800-0.2947 i)  \\
\hline
2 &  &  & & 0.6547-0.4981 i \\
  &  &  & & (0.6539-0.4997 i) \\
\hline  
\end{tabular}
\label{table:Second_set}
\end{table*}

\begin{table*}
\centering
\caption{QN frequencies for $q=0.60,M=1,\mu=0.1$. Massless case in parenthesis}
\begin{tabular}{ccccc}
\hline
$n$ & $l=0$ & $l=1$ & $l=2$ & $l=3$ \\
\hline
0  & 0.1200-0.0952 i & 0.3168-0.0971 i  & 0.5190-0.0976 i  & 0.7227-0.0978 i  \\
   & (0.1181-0.1025 i) & (0.3127-0.0994 i)  & (0.5161-0.0986 i)  & (0.7206-0.0983 i)  \\
\hline
1  &  &  & 0.4994-0.2985 i & 0.7087-0.2962 i  \\
   &  &  & (0.4979-0.3005 i) & (0.7071-0.2975 i)  \\
\hline
2  &  &  &  & 0.6830-0.5023 i \\
   &  &  &  & (0.6822-0.5038 i) \\
\hline   
\end{tabular}
\label{table:Third_set}
\end{table*}

\begin{table*}
\centering
\caption{QN frequencies for $q=0.80,M=1,\mu=0.1$. Massless case in parenthesis}
\begin{tabular}{ccccc}
\hline
$n$ & $l=0$ & $l=1$ & $l=2$ & $l=3$ \\
\hline
0  & 0.1272-0.0969 i & 0.3354-0.0984 i  & 0.5499-0.0988 i  & 0.7661-0.0990 i  \\
   & (0.1252-0.1036 i) & (0.3317-0.1004 i)  & (0.5473-0.0996 i)  & (0.7642-0.0994 i)  \\
\hline   
1  &  &  & 0.5322-0.3014 i & 0.7533-0.2994 i  \\
   &  &  & (0.5307-0.3032 i) & (0.7518-0.3005 i)  \\
\hline
2  &  &  &  & 0.7299-0.5067 i \\
   &  &  &  & (0.7291-0.5080 i) \\
\hline   
\end{tabular}
\label{table:Fourth_set}
\end{table*}

\begin{table*}
\centering
\caption{QN frequencies for $q=1.00,M=1,\mu=0.1$. Massless case in parenthesis}
\begin{tabular}{ccccc}
\hline
$n$ & $l=0$ & $l=1$ & $l=2$ & $l=3$ \\
\hline
0  & 0.1384-0.0979 i & 0.3667-0.0987 i  & 0.6022-0.0991 i  & 0.8394-0.0992 i  \\
   & (0.1364-0.1036 i) & (0.3635-0.1003 i)  & (0.6000-0.0997 i)  & (0.8377-0.0995 i)  \\
\hline
1  &  &  & 0.5874-0.3011 i & 0.8287-0.2995 i  \\
   &  &  & (0.5861-0.3026 i) & (0.8274-0.3004 i)  \\
\hline
2  &  &  &  & 0.8091-0.5052 i \\
   &  &  &  & (0.8083-0.5063 i) \\
\hline   
\end{tabular}
\label{table:Fifth_set}
\end{table*}

\begin{table*}
\centering
\caption{QN frequencies for $q=1.20,M=1,\mu=0.1$. Massless case in parenthesis}
\begin{tabular}{ccccc}
\hline
$n$ & $l=0$ & $l=1$ & $l=2$ & $l=3$ \\
\hline
0  & 0.1812-0.0724 i & 0.4280-0.0894 i  & 0.7083-0.0893 i  & 0.9894-0.0893 i  \\
   & (0.1525-0.0885 i) & (0.4261-0.0902 i)  & (0.7068-0.0896 i)  & (0.9883-0.0894 i)  \\
\hline
1  &  &  & 0.6938-0.2697 i & 0.9793-0.2686 i  \\
   &  &  & (0.6928-0.2704 i) & (0.9783-0.2690 i)  \\
\hline
2  &  &  &  & 0.9593-0.4504 i \\
   &  &  &  & (0.9583-0.4512 i) \\
\hline   
\end{tabular}
\label{table:Sixth_set}
\end{table*}

\begin{table*}
\centering
\caption{QN frequencies for RN and regular black holes for $M=1,\mu=0.1,l=1,n=0$.}
{
\begin{tabular}{ccc}
\hline
$q$ & $\text{RN}$ & $\text{Regular BH}$ \\
\hline
0.20  & 0.299354-0.095294 i &  0.299347-0.095295 i  \\
\hline
0.40  & 0.305624-0.095966 i & 0.305500-0.095993 i  \\
\hline
0.60  & 0.317575-0.096893 i & 0.316803-0.097092 i  \\
\hline
0.80  & 0.338881-0.097145 i & 0.335447-0.098359 i  \\
\hline
0.95  & 0.367320-0.093187 i & 0.357181-0.098879 i  \\
\hline
0.99  & 0.378067-0.089483 i & 0.364714-0.098785 i \\ 
\hline
\end{tabular} 
\label{table:Penultimo_set}
}
\end{table*}

\begin{table*}
\centering
\caption{QN frequencies for RN and regular black holes for $M=1,\mu=0.1,l=2,n=0$.}
{
\begin{tabular}{ccc} 
\hline
$q$ & $\text{RN}$ & $\text{Regular BH}$ \\
\hline
0.20  & 0.490056-0.095905 i &  0.490044-0.095907 i  \\
\hline
0.40  & 0.500432-0.096541 i & 0.500231-0.096570 i  \\
\hline
0.60  & 0.520216-0.097403 i & 0.518963-0.097612 i  \\
\hline
0.80  & 0.555564-0.097572 i & 0.549923-0.098800 i  \\
\hline
0.95  & 0.603642-0.093650 i & 0.586164-0.099245 i  \\
\hline
0.99  & 0.622997-0.089806 i & 0.598769-0.099130 i \\ 
\hline
\end{tabular} 
\label{table:Last_set}
}
\end{table*}

\section{Conclusions}\label{Conclusions}

In this article we have computed the quasinormal modes of four-dimensional charged regular black holes in the presence of Non-Linear Electrodynamical sources. We have studied scalar perturbations using a Schr{\"o}dinger-like equation with the appropriate effective potential, and we have adopted the popular and extensively used WKB approximation of 6th order. Our numerical results are summarized in tables, and we have shown graphically the impact on the spectrum of the electric charge of the black hole as well as of the overtone number and of the quantum number of angular momentum. All modes are found to be stable. A comparison with the RN and other charged black holes is made, and an analytical expression for the QN spectrum in the eikonal limit has been obtained.


\section*{Acknowlegements}

We wish to thank the anonymous reviewer for useful comments and suggestions. The author G.~P. thanks the Funda\c c\~ao para a Ci\^encia e Tecnologia (FCT), Portugal, for the financial support to the Center for Astrophysics and Gravitation-CENTRA, Instituto Superior T\'ecnico, Universidade de Lisboa, through the Grant No. UID/FIS/00099/2013. The author A.~R. acknowledges DI-VRIEA for financial support through Proyecto Postdoctorado 2019 VRIEA-PUCV.


\end{document}